\documentclass[12pt]{article}
\usepackage{axodraw}
\usepackage{epsfig}
\def\e{{\rm e}}
\newcommand{\be}{\begin{equation}}
\newcommand{\ee}{\end{equation}}
\newcommand{\bea}{\begin{eqnarray}}
\newcommand{\eea}{\end{eqnarray}}
\newcommand{\al}{\alpha}

\newcommand{\gm}{\gamma}
\newcommand{\Gm}{\Gamma}
\newcommand{\dl}{\delta}

\newcommand{\ep}{\epsilon}

\newcommand{\dd}{\mbox{d}}

\newcommand{\nn}{\nonumber}

\newcommand{\Li}[2]{{\mbox{Li}}_{#1}\left(#2\right)}

\begin{document}
\parindent=1.5pc

\begin{titlepage}
\rightline{TTP00-24}
\rightline{hep-ph/0011056}
\rightline{November 2000}
\bigskip
\begin{center}
{{\bf
Analytical Result for Dimensionally Regularized \\
Massless Master Non-planar Double Box with One Leg off Shell
} \\
\vglue 5pt
\vglue 1.0cm
{ {\large V.A. Smirnov\footnote{E-mail: smirnov@theory.npi.msu.su.}
} }\\
\baselineskip=14pt
\vspace{2mm}
{\em Nuclear Physics Institute of Moscow State University}\\
{\em Moscow 119899, Russia}
\vglue 0.8cm
{Abstract}}
\end{center}
\vglue 0.3cm
{\rightskip=3pc
 \leftskip=3pc
\noindent The dimensionally regularized massless non-planar double box
{}Feynman diagram with powers of propagators equal to one, one leg
off the mass shell, i.e. with $p_1^2=q^2\neq 0$, and three legs
on shell, $p_i^2=0,\;i=2,3,4$, is analytically calculated for general
values of $q^2$ and the Mandelstam variables $s,t$ and $u$
(not necessarily restricted by the physical condition
$s+t+u=q^2$). An explicit
result is expressed through  (generalized) polylogarithms, up to the
fourth order,  dependent on rational combinations of $q^2,s,t$
and~$u$, and simple finite two- and three fold Mellin--Barnes integrals
of products of gamma functions which are easily numerically evaluated
for arbitrary non-zero values of the arguments.
\vglue 0.8cm}
\end{titlepage}

%\section{Introduction}

{\bf 1.}
Feynman diagrams with four external lines contribute to many important
physical quantities. They are rather complicated mathematical objects
because depend at least on two independent Mandelstam variables.
In the pure massless case with all end-points on shell, i.e.,
$p_i^2=0,\;i=1,2,3,4$, the problem of the analytical evaluation
of two-loop four point diagrams (when the planar and non-planar double
boxes shown in Fig.~1 are mostly complicated),
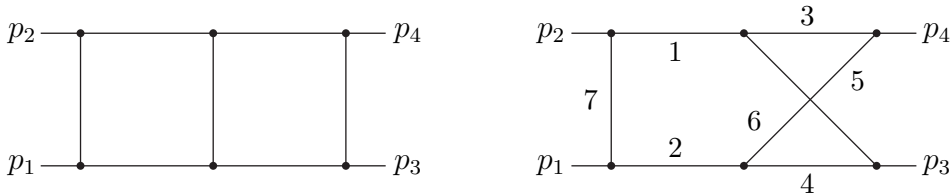
\begin{figure}[hbt]
%\centering
\begin{picture}(400,100)(-50,-20)
%\SetScale{1.3}
\Line(-15,0)(0,0)
\Line(-15,50)(0,50)
\Line(115,0)(100,0)
\Line(115,50)(100,50)
\Line(0,0)(50,0)
\Line(50,0)(100,0)
\Line(100,50)(50,50)
\Line(50,50)(0,50)

\Line(0,50)(0,0)
\Line(50,0)(50,50)
\Line(100,0)(100,50)
%%%%%%%%%%%%%
\Vertex(0,0){1.5}
\Vertex(50,0){1.5}
\Vertex(100,0){1.5}
\Vertex(0,50){1.5}
\Vertex(50,50){1.5}
\Vertex(100,50){1.5}

\Text(-22,0)[]{$p_1$}
\Text(124,0)[]{$p_3$}
\Text(-22,50)[]{$p_2$}
\Text(124,50)[]{$p_4$}

%%%%%%%%%%%%%%%%%%%%%%%%%%%%%%%%%%%%%%%%%%%%%%%%%%%%%%%%%
%%%%%%%%%%%%%%%%%%%%%%%%%%%%%%%%%%%%%%%%%%%%%%%%%%%%%%%%%

\Line(185,0)(200,0)
\Line(185,50)(200,50)
\Line(315,0)(300,0)
\Line(315,50)(300,50)

\Line(200,0)(250,0)
\Line(250,0)(300,0)
\Line(300,50)(250,50)
\Line(250,50)(200,50)

\Line(200,50)(200,0)
\Line(250,0)(300,50)
\Line(300,0)(250,50)

\Vertex(200,0){1.5}
\Vertex(250,0){1.5}
\Vertex(300,0){1.5}
\Vertex(200,50){1.5}
\Vertex(250,50){1.5}
\Vertex(300,50){1.5}

\Text(178,0)[]{$p_1$}
\Text(324,0)[]{$p_3$}
\Text(178,50)[]{$p_2$}
\Text(324,50)[]{$p_4$}

\Text(193,25)[]{\small 7}
\Text(255,17)[]{\small 6}
\Text(294,32)[]{\small 5}
\Text(225,7)[]{\small 2}
\Text(275,-7)[]{\small 4}
\Text(225,43)[]{\small 1}
\Text(275,57)[]{\small 3}
\end{picture}
\caption{The planar and non-planar double boxes}
\end{figure}
in expansion in $\ep=(4-d)/2$ in the framework of dimensional regularization
\cite{dimreg} with the space-time dimension $d$ as a
regularization parameter,
was completely solved during last year in
\cite{K1,SV,Tausk,AGO}. The corresponding analytical algorithms were
quite recently applied to the evaluation of two-loop virtual
corrections to $e^+e^-\to \mu^+\mu^-$ and Bhabba scattering \cite{bhabba}
and quark scattering \cite{Durham}.

In the massless case with one-leg off shell, $p_1^2=q^2\neq 0$,
$p_i^2=0,\;i=2,3,4$, which is relevant to the process $e^+e^-\to 3$jets
(see, e.g., \cite{3jets}),
the planar double box diagram with powers of propagators equal to one
has been
analytically calculated in \cite{S2} for general
values of $q^2$ and the Mandelstam variables $s$ and $t$. An explicit
result is expressed through  (generalized) polylogarithms, up to the
fourth order,  dependent on rational combinations of $q^2,s$
and~$t$, and a one-dimensional integral with a simple integrand
consisting of logarithms and dilogarithms. This result
represented through so-called two-dimensional harmonical polylogarithms
was confirmed in \cite{GR2} where the method of differential equations
\cite{DE,GR1} was applied and some other planar master Feynman diagrams
with one leg off shell were also evaluated.

There are two different non-planar Feynman diagrams (see Fig.~1)
with all powers of propagators equal to one: when $p_1^2\neq 0$
(or $p_2^2\neq 0$) and when $p_3^2\neq 0$ (or $p_4^2\neq 0$).
The purpose of this paper is to
analytically evaluate the first of these two master integrals.
As in the pure on-shell case (see \cite{Tausk}) it turns out that
it is natural to consider non-planar double boxes as functions of
$s,t,u=(p_1+p_4)^2$  and $q^2=p_1^2$ not necessarily restricted by the
physical condition $s+t+u=q^2$ which does not simplify the result.
We shall present an explicit result for the considered master non-planar
double box in terms of (generalized)
polylogarithms, up to the fourth order,  dependent on rational combinations of
$q^2,s,t$ and~$u$, and simple finite two- and three fold Mellin--Barnes (MB)
integrals of products of gamma functions which are easily numerically evaluated
for arbitrary non-zero values of the arguments.

To arrive at this result we straightforwardly apply the method of
refs.~\cite{K1,S2}: we start from the
alpha-representation of the double box and, after expanding some
of the involved functions
in MB integrals, arrive at a five-fold
MB integral representation with gamma functions in the integrand.
Then we use a standard procedure of taking residues and shifting
contours\footnote{The procedures presented in \cite{K1} and \cite{Tausk}
are a little bit different. In the latter variant, one systematically uses
integration in MB integrals in straight lines along imaginary axes.
We prefer to use the former variant.}
to resolve the structure of singularities in
the parameter of dimensional regularization, $\ep$.
This leads to the appearance of multiple terms where
Laurent expansion in $\ep$ becomes possible and provides our result.

Finally we discuss the problem of systematical evaluation of
general double boxes with one leg off shell and the most general class
of functions that can appear in the results.

{\bf 2.}
The alpha representation of the non-planar double box with one leg off shell,
$p_1^2=q^2 \neq 0$ differs from its on-shell variant (see, e.g.,
eqs.~(4)--(6) of ref.~\cite{Tausk}) by a term depending on $q^2$
inserted into one of the functions involved:
\be
{}F (s,t,u,q^2;\ep) =
-\Gm(3+2\ep)\left(i\pi^{d/2} \right)^2
\int_0^\infty \dd\al_1 \ldots\int_0^\infty\dd\al_7
\dl\left( \sum \al_i-1\right) D^{1+3\ep}
A^{-3-2\ep} \; ,
\label{alpha}
\ee
where
\bea
D&=&(\al_1+\al_2+\al_7) (\al_3+\al_4+\al_5+\al_6)
+ (\al_3+\al_5) (\al_4+\al_6) \;, \\
A&=& [\al_1\al_2 (\al_3+\al_4+\al_5+\al_6) + \al_1\al_4 \al_5
+\al_2\al_3\al_6](-s)
\nn \\&& \hspace*{-3mm}
+ \al_5\al_6\al_7 (-t)+ \al_3\al_4\al_7 (-u)
+\al_7 [ \al_4\al_5 +\al_2(\al_3+\al_4+\al_5+\al_6)] (-q^2)
 \; .
\eea
%As it is well-known, one can choose
%a sum of an arbitrary subset of $\al_i\,, i=1,\ldots,7$ in
%the argument of the delta function in (\ref{alpha}).

By introducing six times MB representation
\be
\frac{1}{(X+Y)^{\nu}} = \frac{1}{\Gm(\nu)}
\frac{1}{2\pi i}\int_{-i \infty}^{+i \infty} \dd w
\frac{Y^w}{X^{\nu+w}} \Gm(\nu+w) \Gm(-w) \;,
\label{MB}
\ee
in a suitable way, one can take all the parametrical integrals in gamma
functions. Fortunately, one of the MB integrations is then explicitly
performed by use of the first Barnes lemma and we arrive at the
following 5-fold MB integral:
\bea
{}F (s,t,q^2;\ep) &=&
- \frac{\left(i\pi^{d/2} \right)^2
\Gm(-\ep)^2}{\Gm(-1-3\ep)\Gm(-2\ep)(-s)^{3+2\ep}}
\frac{1}{(2\pi i)^5} \int
\dd v \dd w_1 \dd w_2 \dd z_1 \dd z_2 \left(\frac{q^2}{s} \right)^v
\nn \\ &&  \hspace*{-27mm}\times
\left(\frac{t}{s} \right)^{w_1}
\left(\frac{u}{s} \right)^{w_2}
%\nn \\ &&  \hspace*{-27mm}\times
\frac{\Gm(3+2\ep+v+w_1+w_2+z_1)\Gm(3+2\ep+w_1+w_2+z_1+z_2)
}{\Gm(3 + 2\ep + w_1 + w_2 +z_1)
\Gm(2+w_1+w_2+z_1+z_2)^2}
\nn \\ &&  \hspace*{-27mm}\times
\Gm(2+\ep+w_1+w_2+z_1+z_2) \Gm(1 + v + w_1 + w_2)
\Gm(1 + w_1 + z_1)
\Gm(1 + w_2 + z_1)
\nn \\ &&  \hspace*{-27mm}\times
\Gm(1 + w_1 + z_2) \Gm(1 + w_2 + z_2)
\Gm(-2 - 2 \ep - v - w_1 - w_2 - z_1)
\nn \\ &&  \hspace*{-27mm}\times
\Gm(-2 - 2 \ep  - w_1 - w_2 - z_2)
\Gm(-v) \Gm(-w_1) \Gm(-w_2)\Gm(-z_1)\Gm(-z_2)
\, .
\label{5MB}
\eea
It differs from its analog for $q^2=0$ (see eqs.~(8)--(10) of
ref.~\cite{Tausk})
by the additional integration in $v$. This variable enters only four
gamma functions in the integrand. When taking minus residue of the
integrand in $v$ one reproduces the corresponding 4-fold MB integral
in \cite{Tausk}.
Integral (\ref{5MB}) is evaluated in expansion in $\ep$, up to a finite
part, by resolving singularities in $\ep$  in the same style as in
\cite{S2}. We obtain and confirm as a by-product the on-shell
result of ref.~\cite{Tausk}.

To present the final result
let us turn to the variables  $x=s/q^2$, $y=t/q^2$ and  $z=u/q^2$:
\be
{}F (s,t,u,q^2;\ep) =
\frac{\left(i\pi^{d/2}
\e^{-\gm_{\rm E}\ep} \right)^2 }{s^2 t u (-q^2)^{2\ep-1}}
 \sum_{i=0}^4  \frac{ f_i (x,y,z)}{\ep^i}
+ O(\ep)
\;.
\ee
We obtain
\bea
f_4 (x,y,z) &=&\frac{1}{4} (1 - x - y z) - \frac{3}{8}( z + y)\;,
\label{ResultEp4}
\\  % \hspace*{-10mm}
f_3 (x,y,z) &=&
-\frac{1}{4}  \left[
        3(1 + x + y + z) + (2 - 2 x - 3 y - 3 z - 2 y z) l_x
\right.
\nn \\ &&  %\hspace*{-10mm}
\left.
+ (2 - 2 x + y - 3 z + 2 y z) l_y
      + (2 - 2 x - 3 y + z + 2 y z) l_z
    \right]
\;,
\label{ResultEp3}
\\
f_2 (x,y,z) &=&
3 (x - 1) \Li{2}{x} + 2 (1 - y) z \Li{2}{y} +
  2 y (1 - z) \Li{2}{z}
\nn \\ &&  \hspace*{-27mm}
  + \frac{1}{4}\left[ (2 - 2 x - 3 y - 3 z - 2 y z) l_x  ^2
        + (2 - 2 x + y - 3 z + 2 y z) l_y  ^2
\right.
\nn \\ &&  \hspace*{-27mm}
\left.
+ (2 - 2 x - 3 y + z + 2 y z) l_z  ^2 \right]
  - 3 (1 - x) \bar{l}_x   l_x
  + 2 z(1 - y) \bar{l}_y   l_y
  + 2 y (1 - z) \bar{l}_z   l_z
\nn \\ &&  \hspace*{-27mm}
  + \frac{1}{2}\left[
      (2 - 2 x + y - 3 z + 2 y z) l_x   l_y
        + (2 - 2 x + z - 3y + 2 y z) l_x   l_z
\right.
 \nn \\ &&  \hspace*{-27mm}
\left.
        + (2 - 2 x + y + z - 2 y z) l_y   l_z  \right]
    + \frac{3}{2} \left[
      (1 - x + y + z) l_x   + (1 + x - y + z) l_y
\right.
\nn \\ &&  \hspace*{-27mm}
\left.
      + (1 + x + y - z) l_z   \right]
%\nn \\ &&  \hspace*{-27mm}
  + \frac{\pi^2}{48}(30(1 - x) + 11(y + z) + 26 y z)
  + 3(1 + x + y + z)
\label{ResultEp2}
\;, \\
f_1 (x,y,z) &=&
3(1 - x) (2 \Li{3}{x} + 5  \Li{3}{1-x})
  + 2 (2 - 2 x + 4 y - z - 3 y z) \Li{3}{y}
\nn \\ &&  \hspace*{-27mm}
  - 3 (1 - y) z \Li{3}{1-y}
  + 2 (2 - 2 x - y + 4 z - 3 y z) \Li{3}{z}
  - 3 y (1 - z) \Li{3}{1-z}
 \nn \\ &&  \hspace*{-27mm}
 -3 (1 - x) (5 \bar{l}_x   - 2 (1 + l_y   + l_z  )) \Li{2}{x}
  - 15 (1 - x) \bar{l}_x   \Li{2}{1-x}
 \nn \\ &&  \hspace*{-27mm}
  + (
(1 - y)(6 - 4  z l_x + 3 z \bar{l}_y + 4  z l_z)
%%%  6 - 6 y - 4 (1 - y) z l_x
%%%  + 3 (1 - y) z \bar{l}_y   + 4 (1 - y) z l_z
- (2 - 2 x + 4 y + z - 5 y z) l_y
   ) \Li{2}{y}
\nn \\ &&  \hspace*{-27mm}
+(
  (1-z) (6- 4 y l_x + 3 y \bar{l}_z + 4  y l_y)
%%%  6 - 6 z - 4 y (1 - z) l_x   + 4 y (1 - z) l_y
%%%        + 3 y (1 - z) \bar{l}_z
- (2 - 2 x + y + 4 z - 5 y z) l_z
      ) \Li{2}{z}
\nn \\ &&  \hspace*{-27mm}
+ 3 (1 - y) z \bar{l}_y   \Li{2}{1-y}
+ 3 y (1 - z) \bar{l}_z   \Li{2}{1-z}
\nn \\ &&  \hspace*{-27mm}
  -\frac{1}{6} (2 - 2 x - 3 y - 3 z - 2 y z) l_x  ^3
  + 3 (1 - x) \bar{l}_x   l_x  ^2 - 15 (1 - x) \bar{l}_x  ^2 l_x
\nn \\ &&  \hspace*{-27mm}
     - \frac{1}{6} (2 - 2 x + y - 3 z + 2 y z) l_y  ^3
  - 2 (1 - y) z \bar{l}_y   l_y  ^2
  + 3 (1 - y) z \bar{l}_y  ^2 l_y
\nn \\ &&  \hspace*{-27mm}
-\frac{1}{6}(2 - 2 x - 3 y + z + 2 y z) l_z  ^3
  - 2 y (1 - z) \bar{l}_z   l_z  ^2
  + 3 y (1 - z) \bar{l}_z  ^2 l_z
\nn \\ &&  \hspace*{-27mm}
  -\frac{1}{2}\left[ (2 - 2 x + y - 3 z + 2 y z) l_x   l_y   (l_x   + l_y  )
        + (2 - 2 x - 3 y + z + 2 y z) l_x   l_z  (l_x   + l_z  )
\right.
 \nn \\ &&  \hspace*{-27mm}
\left.
+ (2 - 2 x + y + z - 2 y z) l_y   l_z  (l_y   + l_z  )
      \right]
  - (2 - 2 x + y + z - 2 y z) l_x   l_y   l_z
\nn \\ &&  \hspace*{-27mm}
  + 6 (1 - x) \bar{l}_x   l_x   (l_y   + l_z  )
  + 4 (1 - y) z \bar{l}_y   l_y   (l_z   - l_x   )
  - 4 y (1 - z) \bar{l}_z   l_z  (l_x   - l_y   )
\nn \\ &&  \hspace*{-27mm}
  -\frac{3}{2} \left[(1 - x + y + z) l_x  ^2 - 4(1 - x) \bar{l}_x   l_x
        + (1 + x - y + z) l_y  ^2
- 4 (1 - y) \bar{l}_y   l_y
\right.
\nn \\ &&  \hspace*{-27mm}
\left.
+ (1 + x + y - z) l_z  ^2 - 4 (1 - z) \bar{l}_z   l_z  \right]
\nn \\ &&  \hspace*{-27mm}
  - 3 ((1 - x - y + z) l_x   l_y
        + (1 + x - y - z) l_y   l_z
        + (1 - x + y - z) l_x   l_z  )
\nn \\ &&  \hspace*{-27mm}
- \frac{\pi^2}{24}\left[(30 - 30 x + 11 y + 11 z + 26 y z) l_x
+ (30 - 30 x + 3 y + 11 z - 14 y z)l_y
\right.
\nn \\ &&  \hspace*{-27mm}
- 60 (1 - x) \bar{l}_x
%\nn \\ &&  \hspace*{-27mm}
\left.
+ 12 (1 - y) z \bar{l}_y
        + (30 - 30 x + 11 y + 3 z - 14 y z) l_z
        + 12 y (1 - z) \bar{l}_z
      \right]
\nn \\ &&  \hspace*{-27mm}
  - 6 \left[(1 - x + y + z) l_x   + (1 + x - y + z) l_y
  + (1 + x + y - z) l_z \right]
- 12 (1 + x + y + z)
\nn \\ &&  \hspace*{-27mm}
    -\frac{\zeta(3)}{6} \left[52(1 - x) - 63 (y + z) - 40 y z\right]
%\nn \\ &&  \hspace*{-27mm}
  - \frac{\pi^2}{8} (11 - 5 (x + y + z))
\;,
\label{ResultEp1}
\eea
where $l_a =\ln a$ and $\bar{l}_a =\ln(1-a)$ for $a=x,y,z$.
Moreover $\Li{a}{z}$ is the polylogarithm \cite{Lewin} and
(in the next formula)
\be
\label{Sab}
  S_{a,b}(z) = \frac{(-1)^{a+b-1}}{(a-1)! b!}
    \int_0^1 \frac{\ln^{a-1}(t)\ln^b(1-zt)}{t} \dd t \;
\ee
the generalized polylogarithm \cite{GenPolyLog}.

The finite part $f_0 = \bar{f}_0+\tilde{f}_0$ consists of two pieces:
\bea
\bar{f}_0(x,y,z) &=&
30 (1 - x)S_{2,2}(x) - 2 (1 - y) (6 - 7 z)S_{2,2}(y) -
  2 (1 - z) (6 - 7 y)S_{2,2}(z)
\nn \\ &&  \hspace*{-27mm}
+ 6 (x - y z) \Li{4}{y z/x}  - 12 (1 - x)\Li{4}{x}
\nn \\ &&  \hspace*{-27mm}
  - 2(3 - 6 x - 3 y - z + 4 y z)\Li{4}{y} -
  2 (3 - 3 x - y - 3 z + 4 y z)\Li{4}{z}
\nn \\ &&  \hspace*{-27mm}
+ 69 (1 - x)\Li{4}{1-x} - 3 (1 - y) z \Li{4}{1-y} -
  3 y (1 - z)\Li{4}{1-z}
\nn \\ &&  \hspace*{-27mm}
  -\frac{1}{2} \left[(14 + 10 y - 7 z + y z)\Li{2}{y}^2
  + (14 - 2 x - 7 y + 10 z + y z)\Li{2}{z}^2\right]
\nn \\ &&  \hspace*{-27mm}
-6(1 - x)\left[(2 - 5 \bar{l}_x  + 2 l_y  + 2 l_z )\Li{3}{x}
+ (4 + 5 l_y  + 5 l_z ) \Li{3}{1-x}\right]
\nn \\ &&  \hspace*{-27mm}
- \left[4(2 + x + 4 y - z - 3 y z)l_x  + 2(1 + 5 y - 5 y z) l_y  +
        4(2 - x + 4 y + z + 3 y z)l_z
\right.
 \nn \\ &&  \hspace*{-27mm}
\left.
    + 2 (1 - y) (6 - 7 z)\bar{l}_y  +
        12 (1 - y)\right]\Li{3}{y}
\nn \\ &&  \hspace*{-27mm}
- (6 (1 - y) (4 - l_x  z + l_z  z)
  + (20 + 4 y - 17 z + 11 y z)l_y )\Li{3}{1-y}
\nn \\ &&  \hspace*{-27mm}
- \left[4 (2 + x - y + 4 z - 3 y z) l_x
+ 4 (2 - 2 x + y + 4 z + 3 y z)l_y  +
        2(1 - x + 5 z - 5 y z)l_z
\right.
 \nn \\ &&  \hspace*{-27mm}
\left.
        + 2 (6 - 7 y) (1 - z)\bar{l}_z  + 12 (1 - z)
      \right]\Li{3}{z}
-\left[6 (4 - l_x  y + l_y  y) (1 - z)
\right.
\nn \\ &&  \hspace*{-27mm}
\left.
    + (20 - 2 x - 17 y + 4 z + 11 y z)l_z \right]\Li{3}{1-z}
%\nn \\ &&  \hspace*{-27mm}
  + 6 (l_x  - l_y  - l_z ) (x - y z) \Li{3}{y z/x}
\nn \\ &&  \hspace*{-27mm}
  - \frac{3}{2}(1 - x) \left[4( l_y  + l_z  + 1)^2 + \pi^2 + 12\right]\Li{2}{x}
\nn \\ &&  \hspace*{-27mm}
+\left[4 (1 - y) z (l_x  - l_z )^2
          + (2 - 2 x + 4 y + z - 5 y z)l_y ^2
          + 2(2 + x + 4 y + z - 5 y z)l_x  l_y
\right.
 \nn \\ &&  \hspace*{-27mm}
            + 2(2 - x + 4 y - z + 5 y z) l_y  l_z
      - \bar{l}_y  l_y  (14 + 10 y - 7 z + y z)
      - 12  (1 - y)(l_x  + l_z )
\nn \\ &&  \hspace*{-27mm}
\left.
          + \frac{\pi^2}{6}(22 y + 15 z - 21 y z) - 24(1 - y)
      + \frac{\pi^2}{3}\right]\Li{2}{y}
\nn \\ &&  \hspace*{-27mm}
  + \left[4 y (1 - z)(l_x  - l_y )^2 + (2 - 2 x + y + 4 z - 5 y z)l_z ^2
        + 2 (2 + x + y + 4 z - 5 y z)l_x  l_z
\right.
 \nn \\ &&  \hspace*{-27mm}
+ 2(2 - 2 x - y + 4 z + 5 y z) l_y  l_z
- (14 - 2 x - 7 y + 10 z + y z)\bar{l}_z  l_z
        - 12 (1 - z)(l_x  + l_y )
\nn \\ &&  \hspace*{-27mm}
\left.
        - \frac{\pi^2}{6}(2 x - 15 y - 22 z + 21 y z) - 24(1 - z)
    + \frac{\pi^2}{3}\right]\Li{2}{z}
\nn \\ &&  \hspace*{-27mm}
      +(x - y z)(3(l_x  - l_y  - l_z )^2 + 4 \pi^2) \Li{2}{y z/x}
\nn \\ &&  \hspace*{-27mm}
+\frac{1}{12}(2 - 2 x - 3 y - 3 z + y z)l_x ^4 - 2 (1 - x)l_x ^3 \bar{l}_x
  + \frac{15}{2}(1 - x)l_x ^2 \bar{l}_x ^2
\nn \\ &&  \hspace*{-27mm}
  + \frac{1}{12}(2 - 2 x + y - 3 z + 5 y z)l_y ^4
  + \frac{4}{3}(1 - y) z l_y ^3 \bar{l}_y
  - (10 + 2 y - 7 z + 4 y z)l_y ^2 \bar{l}_y ^2
\nn \\ &&  \hspace*{-27mm}
  + \frac{1}{12}(2 - 2 x - 3 y + z + 5 y z)l_z ^4
  + \frac{4}{3} y (1 - z)l_z ^3 \bar{l}_z
  - (10 - x - 7 y + 2 z + 4 y z)l_z ^2 \bar{l}_z ^2
\nn \\ &&  \hspace*{-27mm}
  + \frac{1}{3} (2 - 2 x + y - 3 z - y z)l_x  l_y (l_x ^2 + l_y ^2)
%\nn \\ &&  \hspace*{-27mm}
  + \frac{1}{2} (2 - 2 x + y - 3 z + 5 y z)l_x ^2l_y ^2
\nn \\ &&  \hspace*{-27mm}
  + \frac{1}{3} (2 - 2 x - 3 y + z - y z)l_x  l_z (l_x ^2 + l_z ^2)
  + \frac{1}{2} (2 - 2 x - 3 y + z + 5 y z)l_x ^2l_z ^2
\nn \\ &&  \hspace*{-27mm}
  + \frac{1}{3} (2 - 2 x + y + z + y z)l_y  l_z (l_y ^2 + l_z ^2)
  + \frac{1}{2} (2 - 2 x + y + z + 9 y z)l_y ^2l_z ^2
\nn \\ &&  \hspace*{-27mm}
  - 6 (1 - x)\bar{l}_x  l_x  (l_y  + l_z ) (l_x  + l_y  + l_z )
\nn \\ &&  \hspace*{-27mm}
  + 4 (1 - y) z \bar{l}_y  l_y  (l_x  - l_z ) (l_x  + l_y  - l_z )
  + 4 y (1 - z)\bar{l}_z  (l_x  - l_y ) l_z  (l_x  - l_y  + l_z )
\nn \\ &&  \hspace*{-27mm}
+ \left[
        (2 - 2 x + y + z + y z)l_x  + (2 - 2 x + y + z - 5 y z)(l_y  +
                l_z ) \right]l_x  l_y  l_z
\nn \\ &&  \hspace*{-27mm}
  - (x - y z)(l_x  - l_y  - l_z )^3\ln(1-y z/x)
\nn \\ &&  \hspace*{-27mm}
  + (1 - x + y + z) l_x ^3 - 6 (1 - x)l_x ^2 \bar{l}_x
  + (1 + x - y + z) l_y ^3 - 6 (1 - y)l_y ^2 \bar{l}_y
\nn \\ &&  \hspace*{-27mm}
  + (1 + x + y - z) l_z ^3 - 6 (1 - z)l_z ^2 \bar{l}_z
  +  3 ((1 - x - y + z)l_x  l_y  (l_x  + l_y )
\nn \\ &&  \hspace*{-27mm}
  + (1 - x + y - z)l_x  l_z  (l_x  + l_z )
        + (1 + x - y - z) l_y  l_z  (l_y  + l_z ))
  + 6 (1 - x - y - z) l_x  l_y  l_z
\nn \\ &&  \hspace*{-27mm}
  - 12 \left[
      l_x  \bar{l}_x  (l_y  + l_z ) (1 - x) +
        l_y  \bar{l}_y (l_x  + l_z ) (1 - y)
    + (1 - z)(l_x  + l_y ) l_z  \bar{l}_z \right]
\nn \\ &&  \hspace*{-27mm}
    + \pi^2
\left\{ \frac{1}{24} (30 - 30 x + 11 z + y (11 + 74 z)) l_x ^2
      - \frac{13}{2} (1 - x) l_x   \bar{l}_x
\right.
 \nn \\ &&  \hspace*{-27mm}
      + \frac{1}{24}(30 - 30 x + 3 y + 11 z + 98 y z) l_y ^2
      +  \frac{1}{6}(14 + y (10 - 7 z) + z) l_y  \bar{l}_y
 \nn \\ &&  \hspace*{-27mm}
      +  \frac{1}{24}(30 - 30 x + 11 y + 3 z + 98 y z) l_z ^2
      +  \frac{1}{6}(14 - 2 x + y + 10 z - 7 y z)l_z  \bar{l}_z
\nn \\ &&  \hspace*{-27mm}
      +  \frac{1}{12}\left[
          (30(1 - x) + 3 y + 11 z - 62 y z)l_x  l_y  + (30(1 - x) + 11 y + 3 z -
                  62 y z)l_x  l_z
\right.
 \nn \\ &&  \hspace*{-27mm}
\left. \left.
            + (30(1 - x) + 3 y + 3 z + 50 y z)l_y  l_z \right]
      - 4 (l_x  - l_y  - l_z ) (x - y z) \ln(1-y z/x)
    \right\}
\nn \\ &&  \hspace*{-27mm}
  + 6\left[
      (1 - x + y + z)l_x ^2 - 4 (1 - x)l_x  \bar{l}_x
        + (1 + x - y + z) l_y ^2 - 4 (1 - y)l_y  \bar{l}_y
\right.
 \nn \\ &&  \hspace*{-27mm}
 \left.
        + (1 + x + y - z) l_z ^2 - 4 (1 - z) l_z  \bar{l}_z
      \right]
 \nn \\ &&  \hspace*{-27mm}
  + 12 (
      (1 - x - y + z)l_x  l_y  + (1 - x + y - z)l_x  l_z
        + (1 + x - y - z)l_y  l_z
      )
 \nn \\ &&  \hspace*{-27mm}
+\zeta(3) \left[
      - \frac{1}{3}(38(1 - x) + 63 (y + z) + 40 y z) l_x
      - 30 (1 - x)\bar{l}_x
 \right.
 \nn \\ &&  \hspace*{-27mm}
        + \frac{4}{3}(28 - 13 x + 14 y - 24 z + 10 y z) l_y  +
        2 (1 - y) (6 - 7 z) \bar{l}_y
\nn \\ &&  \hspace*{-27mm}
\left.
        + \frac{2}{3}(56 - 29 x - 48 y + 28 z + 20 y z) l_z  +
        2 (6 - 7 y) (1 - z) \bar{l}_z
      \right]
\nn \\ &&  \hspace*{-27mm}
  + \frac{\pi^2}{4} \left[ (11(1 - x) - 5( y + z))l_x  + (11(1 - y) -
              5 (x + z))l_y  + (11(1 - z) - 5 (x + y))l_z \right]
\nn \\ &&  \hspace*{-27mm}
  + 24 ( (1 - x + y + z)l_x  + (1 + x - y + z)l_y  + (1 + x + y - z)l_z )
\nn \\ &&  \hspace*{-27mm}
  + 48(1 + x + y + z) + \frac{\pi^2}{2} (11 - 5( x + y + z))
  + 4 (17 - x - y - z) \zeta(3)
\nn \\ &&  \hspace*{-27mm}
  + \frac{\pi^4}{2880} (894(1 - x) + 395 (y + z) + 10034 y z)
\label{ResultEp0}
\eea
and the following two- and three-fold MB integrals:
\bea
\tilde{f}_0(x,y,z) &=&
-\frac{6}{(2\pi i)^3}\int \dd v \dd w_1 \dd w_2 \;
x^{-1 - v - w_1 - w_2} y^{1+w_1} z^{1+w_2}
\Gm(3+v+w_1+w_2)
\nn \\ &&  %\hspace*{-20mm}
\times
\Gm(1+v+w_1+w_2) \Gm(1+w_1) \Gm(-w_1) \Gm(-1-w_1) \Gm(-1-v-w_1)
\nn \\ &&  %\hspace*{-20mm}
\times \Gm(1+w_2) \Gm(-w_2)\Gm(-1-w_2) \Gm(-1-v-w_2)
\nn \\ &&  %\hspace*{-15mm}
+ \frac{6}{(2\pi i)^2}\int \dd v \dd w\;
x^{-w} (y^{1 + w} z^{-v} + z^{1 + w} y^{-v})
\nn \\ &&  %\hspace*{-15mm}
\times \Gm(1 + v) \Gm(v) \Gm(-v)
\Gm(1 + w)^2 \Gm(w) \Gm(-w)^2 \Gm(-1-v-w)
\nn \\ &&  \hspace*{-22mm}
+\frac{6}{(2\pi i)^2}\int \dd v \dd w\;
x^{-v-w} (y^{1 + w} + z^{1 + w})
\frac{\Gm(-v)}{v} \Gm(1+w) \Gm(-1-w)^2 \Gm(1+v+w)
\nn \\ &&  \hspace*{-22mm}
\times \Gm(v+w) \Gm(-v-w)
\left[1 + v + w + v (1 + w) (\psi(1 - v) - \psi(1 + w))
\right]
\nn \\ &&  \hspace*{-22mm}
+ \frac{2}{(2\pi i)^2}\int \dd v \dd w\;
y^{1 + w} z^{-1 - v - w} \left[
    (1 + v + w) w ((1 - v) (v + w) x + v z) - 5 (1 - v) v z
    \right]
\nn \\ &&  \hspace*{-22mm}
\times
\frac{\Gm(-v)}{(1-v) v} \Gm(w) \Gm(-w) \Gm(-1-w)
\Gm(1+v+w) \Gm(v+w)\Gm(-1-v-w) \;.
\label{f0tilde}
\eea
The integration contours in the above 3-fold MB integral are in straight
lines along imaginary axes with $-1<$Re$\,v$,
Re$\,w_{1,2}$, Re$\,(v+w_{1,2})$, Re$\,(v+w_1+w_2)<0$ and, in 2-fold
MB integrals,
with $-1<$Re$\,v$, Re$\,w$, Re$\,(v+w)<0$.

{\bf 3.}
Our result~(\ref{ResultEp4})--(\ref{f0tilde}) is in agreement with the leading
power behaviour when $q^2\to 0$ which is obtained by use of the strategy of
regions  of \cite{BS}. It also agrees  with results based
on numerical integration in the space of alpha parameters \cite{BH}
(where the 1\% accuracy for the $1/\ep$ and $\ep^0$ parts is guaranteed).
The result has been also checked with the value of the double box
at $s=q^2$ which was evaluated by the same technique based on MB integrals.
Note that $F(q^2,q^2,t,u)$ is expressed not only through
(generalized) polylogarithms, as the on-shell dobule box in
\cite{Tausk}, but also through a two-fold MB integral that leads to
a one-parametrical integral of dilogarithms
and logarithms.

The two-fold MB integrals in (\ref{f0tilde}) can be converted into
a form similar to the two-fold MB integrals present in the planar
double box \cite{S2}, or, expressed
through so-called two-dimensional harmonic polylogarithms
\cite{GR2} which  generalize harmonic polylogarithms introduced in
\cite{2dHPL}. The three-fold MB integral in (\ref{f0tilde})
can be written, up to some two-fold MB integrals, through the following
three-fold series
\be
\sum_{n,n_1,n_1=0}^{\infty} \frac{
(n+n_1+1)!(n+n_2+1)!}{n! (n+2)!  (n_1+1)^2 (n_2+1)^2 n_1! n_2!}
A_{n,n_1,n_2} x^n y^{n_1} z^{n_2}\;,
\label{3series}
\ee
where $A_{n,n_1,n_2}$ is a linear combination of
$(\psi(n+n_{1,2}+1)^2 + \psi'(n+n_{1,2}+2))$, \linebreak
$\psi(n+n_{1,2}+2)$, and $\psi(n_{1,2}+1)$.
If we replace $A_{n,n_1,n_2}$ by one the corresponding series
can be summed up with a result in terms of polylogarithms, up
to Li$_3$, depending on certain rational combinations of $x,y,z$.
The presence of $A_{n,n_1,n_2}$ makes the situation much more complicated.
The series can be summed up in the form of a two-parametrical integral of
a cumbersome expression with polylogarithms. It is yet unclear where one of
these parametrical  integrations can be explicitly performed. We therefore
prefer to leave the 3-fold MB integral in the result.
This form can be easily used for numerical evaluation for arbitrary
non-zero values of $x,y$ and $z$.

Keeping in mind the characteristic feature of the non-planar diagrams mentioned
above one can conclude that the presence of four scales
(or, three dimensionless variables) in the problem shows that the
two-dimensional harmonic polylogarithms (introduced in \cite{GR2}
just for the problem of the evaluation of the double boxes)
could be insufficient
to express all the results for the double boxes with one leg off shell,
and one can imagine the necessity of
introducing three-dimensional harmonic polylogarithms.

\vspace{0.5 cm}

{\em Acknowledgments.}
I am grateful to Z.~Kunszt for involving me into this problem
and for kind hospitality during my visit to ETH (Z\"urich) in
April--May 2000 where this work was started.
I am thankful to T.~Binoth and G.~Heinrich for comparison of the
presented result with their results based on numerical integration.
Thanks to A.I.~Davydychev and O.L.~Veretin for useful discussions.
This work was supported by the Volkswagen Foundation, contract
No.~I/73611, and by the Russian Foundation for Basic
Research, project 98--02--16981.


\begin{thebibliography}{99}



\bibitem{dimreg}
G.~'t Hooft and M.~Veltman, {\em Nucl.~Phys.} B44 (1972) 189;
C.G.~Bollini and J.J.~Giambiagi, {\em Nuovo Cim.} 12B (1972) 20.

\bibitem{K1}
V.A.~Smirnov, {\em Phys. Lett.} B460 (1999) 397.

\bibitem{SV}
V.A.~Smirnov and O.L.~Veretin, {\em Phys. Lett.} B566 (2000) 469.

\bibitem{Tausk}
J.B.~Tausk, {\em Phys. Lett.} B469 (1999) 225.

\bibitem{AGO}
C.~Anastasiou, E.W.N.~Glover and C.~Oleari,
{\em Nucl. Phys.} B575 (2000) 416;
C.~Anastasiou, T.~Gehrmann, C.~Oleari, E.~Remiddi
and J.B.~Tausk,
{\em Nucl. Phys.} B580 (2000) 577;
T.~Gehrmann and E.~Remiddi, {\em Nucl. Phys. Proc. Suppl.}
89 (2000) 251;
C.~Anastasiou, J.B.~Tausk and M.E.~Tejeda-Yeomans,
{\em Nucl. Phys. Proc. Suppl.} 89 (2000) 262.


\bibitem{bhabba}
Z. Bern, L. Dixon and A. Ghinculov,
hep-ph/0010075.

\bibitem{Durham}
C. Anastasiou, E.W.N. Glover, C. Oleari and M.E. Tejeda-Yeomans,
hep-ph/0010212; hep-ph/0011094.

\bibitem{3jets}
R.K.~Ellis, D.A.~Ross and A.E.~Terrano,
{\em Nucl. Phys.}  B178 (1981) 421.

\bibitem{S2}
V.A.~Smirnov, {\em Phys. Lett.} B491 (2000) 130.

\bibitem{GR2}
T.~Gehrmann and E.~Remiddi, hep-ph/0008287.

\bibitem{DE}
A.V. Kotikov, {\em Phys. Lett.} B254 (1991) 158;
B259 (1991) 314; B267 (1991) 123;

\bibitem{GR1}
E.~Remiddi, {\em Nuovo Cim.} 110A (1997) 1435;
T.~Gehrmann and E.~Remiddi, {\em Nucl. Phys.} B580 (2000) 485.

\bibitem{Lewin}
L.~Lewin, {\em Polylogarithms and associated functions}
(North Holland, 1981).

\bibitem{GenPolyLog}
K.S. Kolbig, J.A. Mignaco and E. Remiddi, {\em B.I.T} 10 (1970) 38;
K.S. Kolbig, {\em Math. Comp.} 39 (1982) 647;
A.~Devoto and D.W.~ Duke, {\em Riv. Nuovo Cim.} 7 (1984) 1.

\bibitem{BS}
M. Beneke and V.A.~Smirnov, {\em Nucl. Phys.} B522 (1998) 321;
V.A.~Smirnov and E.R. Rakhmetov, {\em Teor. Mat. Fiz.} 120 (1999)
64; V.A.~Smirnov, {\em Phys. Lett.}  B465 (1999) 226.

\bibitem{BH}
T.~Binoth and G.~Heinrich, {\em Nucl. Phys.} B585 (2000) 741.

\bibitem{2dHPL}
E. Remiddi and J.A.M. Vermaseren, {\em Int. J. Mod. Phys.} A15 (2000) 725.

\end{thebibliography}
\end{document}